\documentclass[twocolumn,superscriptaddress,amsmath,amssymb,aps,prl,sort&compress,balancelastpage]{revtex4-1}

\usepackage[utf8]{inputenc}

\usepackage{times,color,amsthm,graphics,graphicx,bm,bbm,dcolumn}
\usepackage{epsfig}
\usepackage{comment}
\usepackage{graphicx}
\usepackage{xcolor}
\usepackage{booktabs}
\usepackage[colorlinks,urlcolor=black,citecolor=blue,linkcolor=black]{hyperref}
\usepackage{units}
\usepackage{multibbl}

\setcitestyle{super}

\begin{document}

\title{Sound emission and annihilations in a programmable quantum vortex collider}

\date{\today}

\author{W. J. Kwon}
\email[E-mail: ] {kwon@lens.unifi.it}
\affiliation{European Laboratory for Nonlinear Spectroscopy (LENS), 50019 Sesto Fiorentino, Italy}
\affiliation{Istituto Nazionale di Ottica del Consiglio Nazionale delle Ricerche (CNR-INO), 50019 Sesto Fiorentino, Italy}
\author{G. Del Pace}
\affiliation{European Laboratory for Nonlinear Spectroscopy (LENS), 50019 Sesto Fiorentino, Italy}
\affiliation{Istituto Nazionale di Ottica del Consiglio Nazionale delle Ricerche (CNR-INO), 50019 Sesto Fiorentino, Italy}
\author{K. Xhani}
\affiliation{European Laboratory for Nonlinear Spectroscopy (LENS), 50019 Sesto Fiorentino, Italy}
\affiliation{Istituto Nazionale di Ottica del Consiglio Nazionale delle Ricerche (CNR-INO), 50019 Sesto Fiorentino, Italy}
\author{L. Galantucci}
\affiliation{Joint Quantum Centre (JQC) Durham-Newcastle, School of~Mathematics,~Statistics~and~Physics, Newcastle University, Newcastle
upon Tyne NE1 7RU, United Kingdom}
\author{A.~Muzi~Falconi}
\affiliation{European Laboratory for Nonlinear Spectroscopy (LENS), 50019 Sesto Fiorentino, Italy}
\affiliation{Istituto Nazionale di Ottica del Consiglio Nazionale delle Ricerche (CNR-INO), 50019 Sesto Fiorentino, Italy}
\author{M. Inguscio}
\affiliation{European Laboratory for Nonlinear Spectroscopy (LENS), 50019 Sesto Fiorentino, Italy}
\affiliation{Istituto Nazionale di Ottica del Consiglio Nazionale delle Ricerche (CNR-INO), 50019 Sesto Fiorentino, Italy}
\affiliation{Department of Engineering, Campus Bio-Medico University of Rome, 00128 Rome, Italy}
\author{F. Scazza}
\altaffiliation[Present address: ]{Department of Physics, University of Trieste, 34127 Trieste, Italy}
\affiliation{European Laboratory for Nonlinear Spectroscopy (LENS), 50019 Sesto Fiorentino, Italy}
\affiliation{Istituto Nazionale di Ottica del Consiglio Nazionale delle Ricerche (CNR-INO), 50019 Sesto Fiorentino, Italy}
\author{G. Roati}
\affiliation{European Laboratory for Nonlinear Spectroscopy (LENS), 50019 Sesto Fiorentino, Italy}
\affiliation{Istituto Nazionale di Ottica del Consiglio Nazionale delle Ricerche (CNR-INO), 50019 Sesto Fiorentino, Italy}

\begin{abstract}
In quantum fluids, the quantisation of circulation forbids the diffusion of a vortex swirling flow seen in classical viscous fluids. Yet, a quantum vortex accelerating in a superfluid may lose its  
energy into acoustic radiation \cite{parker2004,barenghi2005}, in a similar way an electric charge decelerates upon emitting photons. The dissipation of vortex energy underlies central problems in quantum hydrodynamics \cite{TSUBOTA_review2012}, such as the decay of quantum turbulence, highly relevant to systems as varied as neutron stars, superfluid helium and atomic condensates \cite{vinen2002Niemela,Barenghi2014review}. A deep understanding of the elementary mechanisms behind irreversible vortex dynamics has been a goal for decades \cite{Feynman1955,TSUBOTA_review2012}, but it is 
complicated by the shortage of conclusive experimental signatures \cite{vinen2007phys,Vinen2010}. Here, we address this challenge by realising a programmable 
quantum vortex collider in a planar, homogeneous atomic Fermi superfluid with tunable inter-particle interactions. We create on-demand vortex configurations and monitor their evolution, taking advantage of the accessible time and length scales of our ultracold Fermi gas \cite{zwierlein2005vortices,ku2016}. Engineering collisions within and between vortex-antivortex pairs allows us to decouple relaxation of the 
vortex energy due to sound emission and interactions with normal fluid, i.e.~mutual friction. 
We directly visualise how the annihilation of vortex dipoles radiates a sound pulse in the superfluid. 
Further, our few-vortex experiments extending across different superfluid regimes suggest that fermionic quasiparticles localised inside the vortex core contribute significantly to dissipation,
opening the route to exploring new pathways for quantum turbulence decay, vortex by vortex.
\end{abstract}

\maketitle

Quantized vortices are a ubiquitous form of topological excitation in quantum matter. Their motion governs the resistive behaviour of superconductors \cite{Bardeen1965,RMP_supercond}, as well as the emergence of many dissipative collective phenomena in superfluids, ranging from vortex lattices to the quantum turbulence of chaotic vortex tangles \cite{TSUBOTA_review2012}. In any quantum fluid, the superflow circulation around a vortex can only take discrete values, making vorticity intrinsically robust. However, travelling vortices interact with the normal component and experience mutual friction \cite{Hall1956}, which causes dissipation of the superflow. Even in the absence of a normal fluid, it has been proposed that vortex-sound interactions may lead to the transformation of incompressible kinetic energy of the superflow associated with vortices into compressible sound energy, providing the ultimate mechanism behind the decay of quantum turbulence \cite{Nore1997,vinen2001,Vinen_tsubota2003}. 
In three-dimensional (3D) vortices, the emission of sound waves is triggered by the accelerated motion of vortex rings \cite{xhaniprl}  and lines at sufficiently short scales, resulting from vortex reconnections \cite{Leadbeater2001,Ruostekoski2005,Irreversible2020}, and helical Kelvin-wave cascades \cite{vinen2001,Kivotides2001,Vinen_tsubota2003,Kozik2004,Lvov2010}. In a similar manner, accelerating two-dimensional (2D) point vortices may dissipate energy by emitting sound waves, much like an accelerated electric charge radiates electromagnetic waves \cite{parker2004}. In such analogy, which becomes apparent by considering the mathematical description of the 2D Bose superfluid and (2+1)D electrodynamics, vortices and phonons play the roles of electric charges and photons, respectively \cite{popov1973,Ambegaokar1980,Arovas1997,Ao2000}. 

Despite prolonged efforts in superfluid helium \cite{bradley2011BBR,Fonda2014PNAS,superdiffusion2021}, sound-mediated dissipation of vortex energy remains elusive due to the scarcity of convincing experimental proofs \cite{vinen2007phys,Walmsley2014}. 
Another central question is how vortex-sound interactions could be influenced by the fermionic or bosonic nature of superfluids \cite{vinen2002Niemela}. In fermionic superfluids, pair-breaking excitations become energetically accessible, and vortices have intricate structures filled up with quasiparticle bound states even at zero temperature \cite{CAROLI1964,Kopnin1991,MicroscopicVortex2003,Ho2006}. Fermionic quasiparticle states in the vortex core are thought to provide an additional dissipation sink by absorbing the vortex kinetic energy \cite{Silaev2012}, which may in turn compete with dissipation via phonon emission \cite{vinen2002Niemela}. This is in stark contrast to weakly interacting bosonic superfluids, where the empty vortex core is directly connected to the vanishing order parameter. %

In this work, we study the fundamental mechanisms of vortex energy dissipation by realising a versatile 2D vortex collider in homogeneous atomic superfluids. We unveil vortex-sound interactions by observing the conversion of the energy of vortex swirling flow into sound energy during vortex collisions. 
We directly visualise vortices annihilating into sound waves, i.e.~the ultimate outcome of small-scale four-vortex collisions, and find good agreement with theoretical simulations, shedding light on such fundamental process \cite{kwon2014,Johnstone2019}. 
Tuning our system away from the bosonic regime into a superfluid of weakly bound fermion pairs, we find indications of the essential role played by vortex-core-bound fermionic excitations, breaking down a picture based solely on sound-mediated vortex energy dissipation at low temperatures, as well as highly influencing mutual friction.
For our studies, we follow a bottom-up approach reminiscent of other atomic platforms featuring control at the single quantum level, and gain exquisite control of individual 2D vortices to assemble them one-by-one in arbitrary arrangements. Such controllable 2D vortex systems, where vortex line excitations have only a moderate role \cite{Rooney2011}, allow us to efficiently track vortex trajectories and hunt for visual evidence of low-energy acoustic radiation. The building blocks of our collider are individual vortex dipoles, namely self-propelling vortex-antivortex pairs carrying a constant linear fluid momentum. A single dipole is ideal for precisely probing the effect of mutual friction, under whose action it gradually shrinks in size and may eventually self-annihilate. Injecting another dipole triggers a dipole-dipole collision \cite{Ruostekoski2005}, providing a minimalistic, yet effective way to promote vortex acceleration, favouring in turn vortex energy dissipation. Thereby, we progress toward a complete microscopic description of the complex dissipative dynamics of both single and colliding vortex-antivortex pairs, which is at the heart of the hydrodynamic relaxation of nonequilibrium states in bosonic and fermionic quantum fluids \cite{zwierlein2005vortices,Vinen2010,TSUBOTA_review2012,Barenghi2014review,ku2016}.

\begin{figure}[t!]
\centering
\vspace{6pt}
\includegraphics[width=0.96\columnwidth]{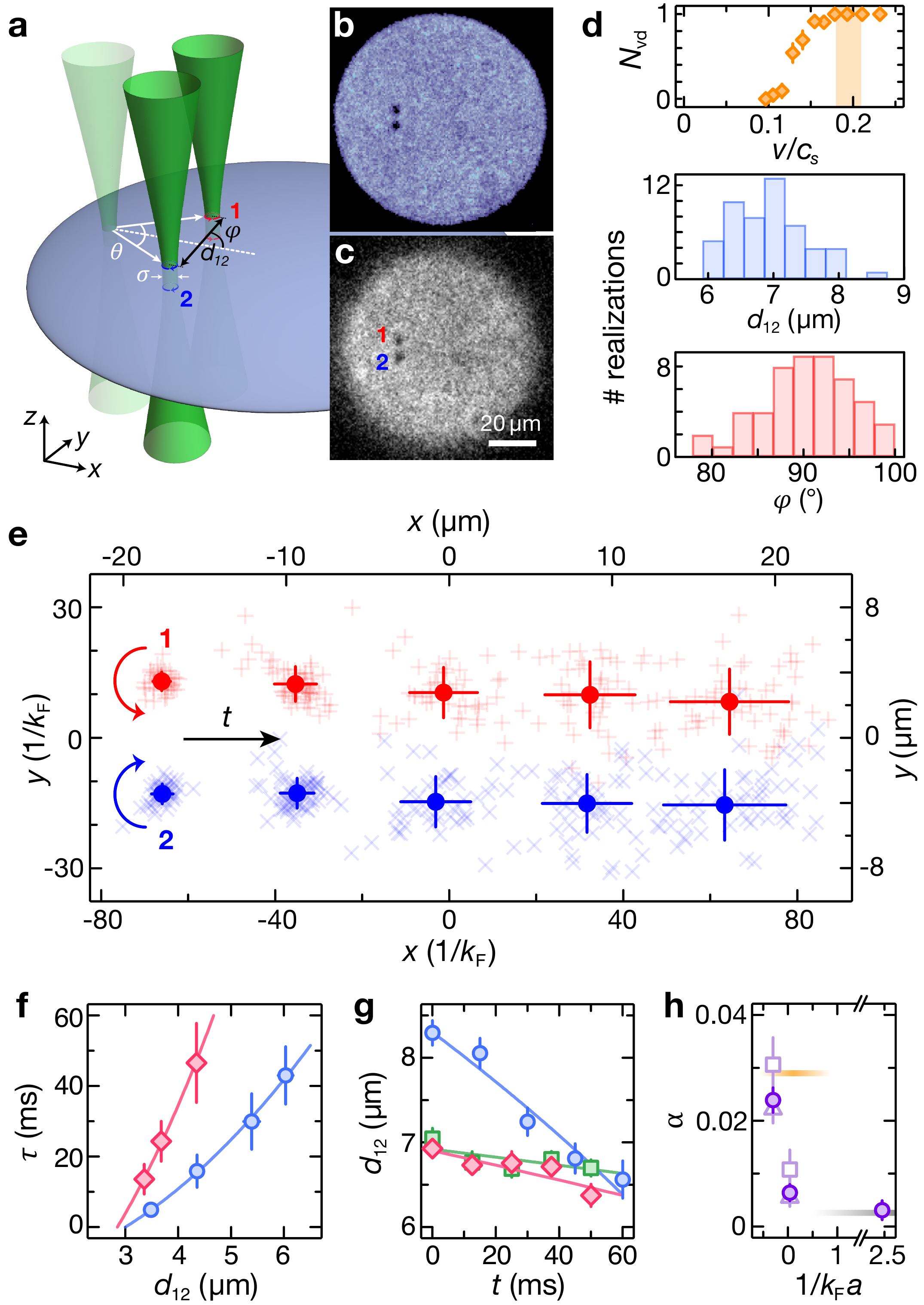}
\vspace*{5pt}
\caption{
\textbf{Deterministic generation and frictional dissipation  
of single vortex dipoles.}
\textbf{a}, Illustration of the dipole generation protocol: we move two obstacles with velocity $v$ and relative angle $\theta$ through a thin, homogeneous superfluid, pinning a vortex-antivortex pair.
\textbf{b}, In-situ image of the UFG sample with obstacles in their final positions. \textbf{c}, Time-of-flight image (Methods) of the UFG acquired after obstacles have been removed, revealing a single vortex-antivortex pair.
\textbf{d}, Properties of generated vortex dipoles in UFGs for $\theta \simeq 20^{\circ}$: mean dipole number $N_\text{vd}$, inter-vortex separation $d_{12}$, and direction $\varphi$.
The histograms are obtained with $v/c_s\simeq 0.2$ and obstacle height $V_0/\mu \simeq 2.2$, where $c_s$ is the sound speed and $\mu$ the pair chemical potential. 
\textbf{e}, Trajectory of single dipoles in UFGs, %
prepared with $d_{12}=6.9 (6)\mu$m and $\varphi=90 (5) ^{\circ}$ (see  \textbf{d}). Data are collected for hold times $0 \leq t \leq 50\,$ms in steps of $12.5$\,ms. Here, light red $+$ (blue $\times$) signs represent the distribution of vortices (antivortices) detected at each time $t$, while red (blue) circles depict their positions averaged over $\sim 40$ experimental runs. 
\textbf{f}, Short dipoles half-lifetime $\tau$ measured in BCS (blue circles) and UFG (red diamonds) samples.
\textbf{g}, Time evolution of $d_{12}$ in BCS (blue circles), UFG (red diamonds), and BEC (green squares) samples. 
Solid lines in panels \textbf{f} and \textbf{g} are fits with a DPV model (Methods).
\textbf{h}, Mutual friction coefficient $\alpha$ across the BEC-BCS crossover. Violet circles denote the weighted mean value of $\alpha$ extracted from \textbf{f} (light squares) and \textbf{g} (light triangles). 
The gray and orange horizontal lines denote the values of $\alpha$ obtained in Ref.~\citenum{jackson2009} for a weakly interacting BEC at $T=0.3\,T_c$, and in Ref.~\citenum{holography2020} for a strongly interacting, 2D holographic Bose superfluid at $T=0.45\,T_c$, respectively.
Error bars of \textbf{d} and \textbf{g} indicate the standard error of the mean (Methods), while they represent the standard deviation for \textbf{e}, and standard fitting errors for \textbf{f} and \textbf{h}.}
\label{Fig1}
\end{figure}
 
Our experiment starts with thin, uniform superfluids of paired fermionic ${}^6$Li atoms trapped inside a circular box of $45\,\mu$m radius (Fig.~\ref{Fig1}a,b) at a temperature $T \simeq 0.3(1)\,T_c$, where $T_c$ is the critical temperature of the superfluid transition. Tuning the $s$-wave scattering length $a$ between two different atomic states, we access three different coupling regimes, namely a Bardeen-Cooper-Schrieffer superfluid of fermionic pairs (BCS, $1/k_Fa\simeq-0.31$), a unitary Fermi gas (UFG, $1/k_Fa\simeq0.04$), and a Bose-Einstein condensate of tightly bound molecules (BEC, $1/k_Fa\simeq2.5$). Here, $k_F=(6 \pi^2 n)^{1/3}$ is the Fermi wave vector, estimated from the central sample density $n$ along the $z$ axis.
To deterministically create a vortex dipole in a desired position and propagation direction, we implement the so-called chopstick method \cite{samson2016}, illustrated in Fig.~\ref{Fig1}a. We initially pierce the cloud with a Gaussian repulsive obstacle focused on the sample with $1/e^2$ radius $\sigma=1.3(1)~\mu$m$~\approx2\xi\approx5/k_F$, where $\xi=0.68(2)~\mu$m is the healing length of our molecular BEC and $1/k_F=0.27(1)~\mu$m is the characteristic Fermi length at unitarity. %
We split the initial obstacle into two and simultaneously move them with velocity $v$ by $10\,\mu$m in oblique directions at a tuning angle $\theta$ (Methods). A vortex-antivortex pair of separation $d_{12}$ is created and pinned around the final positions of the obstacles (Fig.~\ref{Fig1}b), appearing as clear holes in the density of the expanding cloud (Fig.~\ref{Fig1}c). When $v$ exceeds a critical value, our preparation generates a single dipole with near-unit probability and reproducible size $d_{12}$ and direction $\varphi$ (Fig.~\ref{Fig1}d),
resulting in well-defined single-dipole trajectories  (Fig.~\ref{Fig1}e).
We control the size of vortex dipoles by adjusting $\theta$, accessing small length scales down to $d_{12}\lesssim 3\,\mu$m, that is a few times the vortex core radius $\xi_v$. %
Given the homogeneity of the sample (Extended Data Fig.~2), we can fully characterise vortex energy dissipation with the single parameter $d_{12}$, linked both to the incompressible kinetic energy of the superflow $E \propto \log(d_{12}/\xi_v)$ and the linear fluid momentum $P \propto d_{12}$ carried by a vortex dipole \cite{Donnelly1991}. For our experimental $d_{12}$ values larger than a few $\xi_v$, we verify that vortex-dipole speed is inversely proportional to $d_{12}$, 
such that faster dipoles have lower $P$ and $E$.

We first conduct single-dipole experiments, observing a tapering of trajectories over the propagation time $t$. The decrease of $d_{12}(t)$ reflects the effective attraction between vortex and antivortex arising from mutual friction. This causes dipoles to shrink down to a critical size $d_c$ prior to self-annihilating, possibly before reaching the boundary of the cloud. 
We extract the dipole half-lifetime $\tau$ as a function of initial $d_{12}$ in UFG and BCS regimes, by measuring the evolution of the mean dipole number $N_\text{vd}(t)$ (Fig.~\ref{Fig1}f, raw data in Extended Data Fig.~3). Data are fitted with a dissipative point-vortex (DPV) model \cite{Hall1956}, including only a longitudinal mutual friction coefficient $\alpha$ (solid lines, Methods). %
From the $x$-intercept of the fitting curves, we obtain $d_c \simeq 10/k_F\simeq 10\,\xi_v$ in both regimes, %
being $\xi_v \approx 1/k_F$ \cite{BulgacPRL2003,levin2006,Ho2006,Strinati2013}. 
Conversely, in bosonic superfluids we do not observe single-dipole self-annihilations during the propagation, even for $d_{12} \simeq 5\xi$, with $\xi \simeq \xi_v$ (Extended Data Fig.~2). This is consistent with predictions \cite{Jones1982,Ruostekoski2005} $d_c \sim 2\xi$.
Such deviating $d_c$ could stem from the complicated vortex core structures and weaker pairing strength of fermionic (UFG and BCS) superfluids. Nonetheless, we cannot exclude effects of a relatively larger thickness along the $z$-direction in these regimes (Methods), which might allow longer vortex lines to bend and increase $d_c$.

To gain further insight on mutual friction, we monitor the $d_{12}(t)$ evolution for $d_{12} > d_c$ in the three different superfluid regimes (Fig.~\ref{Fig1}g). We fit data with the DPV model, and the extracted $\alpha$ from both $d_{12}(t)$ and lifetime measurements are summarised in Fig.~\ref{Fig1}h. As signaled by the shorter $\tau$ and the steeper drop of $d_{12}(t)$, $\alpha$ is found to sharply increase when crossing to the BCS regime, whereas it slowly decreases when going from unitarity to a BEC. 
This observation does not appear compatible with the scattering of normal excitations in the bulk by the moving vortices, since the former are comparably low in all three regimes.
Instead, this may indicate that the complex vortex core structure of fermionic superfluids, filled with unpaired bound levels \cite{Kopnin1991,MicroscopicVortex2003,Ho2006,zwierlein2005vortices,ku2014}, i.e.~Andreev bound states, affects the mutual friction by interacting with bulk normal excitations \cite{Kopnin1991}. Indeed, the rise of $\alpha$ from the UFG to the BCS regime could be explained by the proliferation of bound quasiparticles in the latter \cite{Ho2006}. 
On the other hand, the small value of $\alpha$  measured in our finite-temperature BEC is found in agreement with previous works \cite{jackson2009, moon2015}, and signifies a small effect from thermal atoms occupying the vortex core. %
We also compare our findings with a recent theoretical prediction\cite{holography2020} $\alpha = 0.029$ based on 2D holographic Bose superfluids at strong coupling and $T/T_c=0.45$, somewhat higher than $\alpha \simeq 0.006$ measured at unitarity, deserving further investigations. %

\begin{figure}
\centering
\includegraphics[width=78mm]{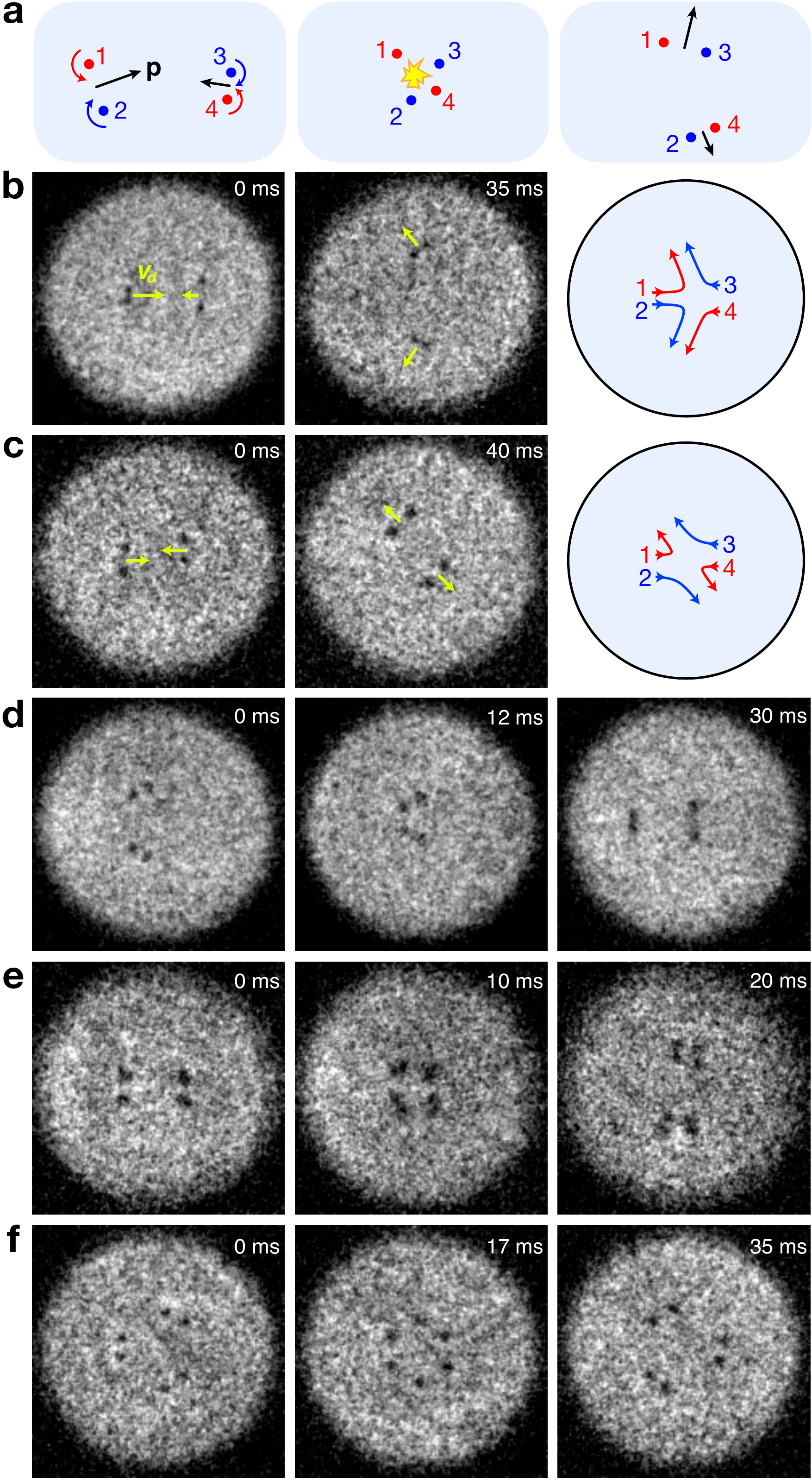}
\caption{\textbf{Arbitrary vortex collisions on-demand.} \textbf{a}, Sketch of the collision between two dipoles (`12' and `34'), during which new pairs (`13' and `24') are formed.
\textbf{b}, Collisions between two dipoles with different sizes and, \textbf{c}, between two dipoles with similar sizes and an impact parameter $b\approx 0.5 \, d_{12}$. 
The right-most panels show the expected trajectories, calculated with the DPV model. %
Arrows in panel \textbf{a} denote the momentum vectors of each dipole, while in panel \textbf{b},\textbf{c} they represent the dipole velocity. \textbf{d}, Oblique collision of two dipoles under a $120^\circ$ angle: 
the resulting dynamics is close to time-reversing that in \textbf{b}. %
\textbf{e}, Head-on collision of two doubly-charged dipoles in UFGs. %
\textbf{f}, Symmetric collision of three dipoles. 
All images are recorded in molecular BECs, except for \textbf{e}, at hold times $t$ noted on top of each panel.}
\label{Fig2}
\end{figure}

\begin{figure*}[htbp]
\centering
\includegraphics[width=170mm]{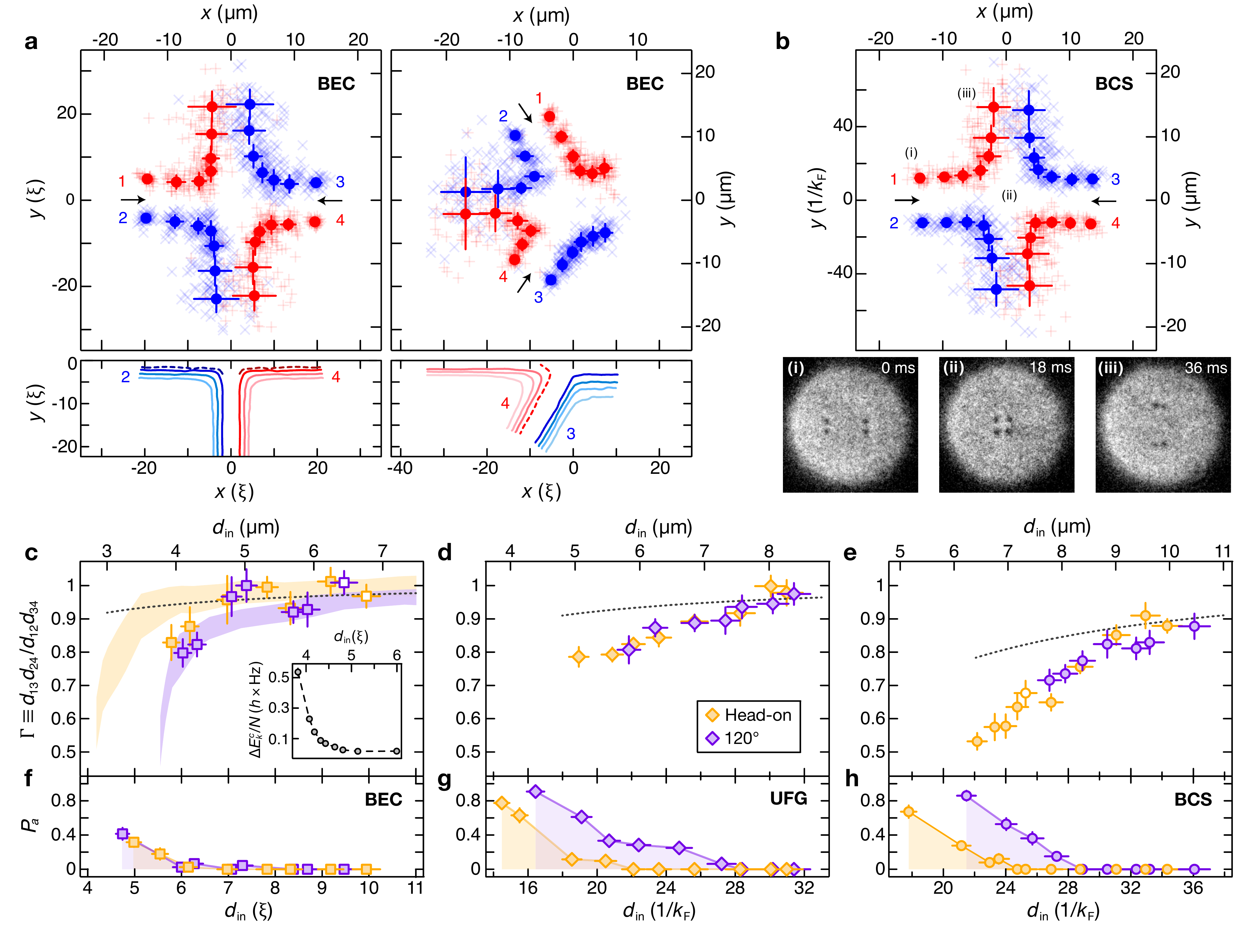}
\vspace*{-14pt}
\caption{\textbf{Dissipation and annihilation in two-dipoles collisions} 
\textbf{a}, Experimental trajectories of head-on (left) and 120$^\circ$ (right) collisions in BECs. 
Transparent $\times$ and $+$ signs indicate the vortex-antivortex positions in single experimental realisations, while full circles and error bars denote the averages and standard deviations over repeated experiments. 
The bottom panels show GPE simulated trajectories for various $d_{in}$. 
Same colours denote equal $d_{in}$, while dashed lines denote the annihilating trajectories.
\textbf{b}, Experimental trajectory of head-on collision in a BCS superfluid, setting $d_{in}$ similar to that of panel \textbf{a}. Vortex images averaged over a few experimental shots are shown for three evolution times (i)-(iii) within the trajectory. 
All experimental trajectories are obtained by $\sim 40$ realisations for each $t$ with the time interval $6$\,ms. \textbf{c-h}, $\Gamma\equiv d_{13}d_{24}/d_{12}d_{34}$ and annihilation probability $P_a$ measured for various $d_{in}$ in the different superfluid regimes. 
Orange (purple) symbols denote the experimental values obtained from head-on ($120^\circ$) collisions. Empty symbols in \textbf{c} and \textbf{e} mark the experiments shown in \textbf{a} and \textbf{b}, respectively. 
Shaded bands and black dotted lines correspond to the predicted values of $\Gamma$ from GPE simulations and DPV model solutions, respectively (Methods).
The inset of \textbf{c} shows the GPE-computed change of compressible kinetic energy $\Delta E_{k}^{c}/N$ throughout a head-on collision, where $N$ is the total number of pairs.
Error bars for \textbf{c-h} indicate the standard errors of the mean over $30-100$ realisations, while they represent the standard deviation of measurements for \textbf{a} and \textbf{b}.}
\label{Fig3}
\end{figure*}

Scaling up our single-dipole control, %
we engineer and demonstrate various paradigmatic vortex collisions \cite{Aioi2011PRX} (Fig.~\ref{Fig2}), serving as the starting point to investigate strong dissipation in binary dipole collisions. 
Figure~\ref{Fig2}a sketches the collision of two arbitrary dipoles `12' and `34': each vortex exchanges its partner and forms a new pair, which then moves away with new momentum. %
We can perform dipole-dipole collisions with controllable dipole sizes (Fig.~\ref{Fig2}b), impact parameter $b$ (Fig.~\ref{Fig2}c) and scattering angle (Fig.~\ref{Fig2}d). When dissipation effects are negligible and $d_{12},d_{34}\gg d_{c}$, the collision dynamics is fully explained by momentum and energy conservation. %
Together with $b$, this determines the scattering angles observed in Figs.~\ref{Fig2}b,c. 
This is evident in Fig.~\ref{Fig2}d, where a time-reversed version of Fig.~\ref{Fig2}b is realised. Further, we extend our deterministic generation to produce doubly-charged dipoles in a UFG (Methods), and investigate their head-on collisions (Fig.~\ref{Fig2}e). %
Notably, during its propagation each doubly-charged vortex splits into a pair of same-circulation vortices due to dynamical instability, forming a charge-2 cluster (Fig.~\ref{Fig2}e, $t=10$\,ms). 
Finally, we exemplify the scalability of our collider by demonstrating a symmetric 6-vortex collision, displayed in Fig.~\ref{Fig2}f. Despite the input and output configurations do not necessarily entail a collision, the exchange of partners between dipoles becomes clear by considering the hexagonal pattern observed at intermediate time. Figure~\ref{Fig2} illustrates the feasibility of programmable manipulation of vortex configurations by tuning the vortex positions, circulations and number on-demand. 

We now focus on the regime of strong dissipation driven by high vortex accelerations within a collision. We systematically study head-on collisions between two symmetrically prepared dipoles with $b\simeq0$ by varying their initial sizes $d_{12}$ and $d_{34}$, and constraining $\langle d_{12}\rangle\approx \langle d_{34}\rangle$ over many realisations. To enhance the effect of acceleration for given initial vortex energy, we also realise oblique collisions at 120$^\circ$ (as in Fig.~\ref{Fig2}d).
Examples of experimental trajectories for both head-on and 120$^\circ$ collisions in BECs are displayed in Fig.~\ref{Fig3}a. Each dipole is prepared at a distance $12.5\,\mu$m $\lesssim L \lesssim 14\,\mu$m from the centre of the cloud, and moves towards the other until the inter-dipole distance becomes comparable to its size. Here, each vortex experiences a strong acceleration as it adjusts its direction. 
Figure~\ref{Fig3}b shows the head-on collision trajectory observed in a BCS superfluid upon setting $d_{12} \simeq d_{34}$ to be similar with that of Fig.~\ref{Fig3}a. In comparison with the bosonic regime, the exiting dipoles have significantly shrunk, signaling that much stronger energy dissipation occurs in fermionic BCS superfluids. %
We quantify the vortex energy loss during a collision by measuring the dipole sizes in the initial and final configurations, when the inter-dipole distance is around $2L$. The ratio between final and initial dipole sizes, $\Gamma\equiv d_{13}d_{24}/d_{12}d_{34}$, directly reflects the change in the kinetic energy of the swirling flow $\Delta E$ caused by the collision. Indeed, the energy of two far-apart dipoles is well approximated by $E\propto \log(d_{12}/\xi_v)+\log(d_{34}/\xi_v)$, hence $\Delta E \propto \log(d_{13}d_{24}/d_{12}d_{34})$. The measured $\Gamma$ are displayed in Figs.~\ref{Fig3}c-e as a function of $d_{in} \equiv\sqrt{\langle d_{12}d_{34}\rangle}$, exhibiting a clear decreasing trend while reducing $d_{in}$ in all superfluid regimes. We compare them with predictions of the DPV model (dotted lines, head-on) obtained by employing the measured friction coefficients $\alpha$ (Fig.~\ref{Fig1}h). 
Whereas for the largest explored $d_{in}$, where $\Gamma \gtrsim 0.9$ in each regime, mutual friction seems sufficient to account for the observed behaviour, the pronounced drop of $\Gamma$ for smaller $d_{in}$ reveals that other sources of dissipation kick in during short-scale dipole collisions. 
Since the finite compressibility of our superfluid can become important when vortices start to overlap, we expect the enhanced incompressible kinetic energy loss at short $d_{in}$ scales to be associated with sound-like density excitations. 

We compare our collision measurements in the BEC regime with Gross–Pitaevskii equation (GPE) simulation results.
From numerical trajectories like those shown in Fig.~\ref{Fig3}a, we extract the trends of $\Gamma$ (Methods)
and plot them in Fig.~\ref{Fig3}c as shaded bands.
We find clear monotonic behaviours for both
head-on and 120$^\circ$ collisions, matching the observed 
reduction of $\Gamma$ for small $d_{in}$. We compute also the compressible kinetic energy variation $\Delta E_{k}^{c}$ throughout the head-on collision (Fig.~\ref{Fig3}a, inset), which evidences how part of the incompressible kinetic energy is indeed converted to compressible sound energy. To elucidate the origin of the sharp decrease of $\Gamma$, it is instructive to estimate the total acoustic energy radiated by an accelerating vortex in analogy with the Larmor formula for the power radiated by an accelerating charge. Assuming symmetric head-on collisions and negligible vortex cores, this gives approximately \cite{Ruostekoski2005} $\Delta E_{k}^{c} \propto 1/d_{in}^{4}$. This qualitatively explains the steep behaviour observed in BEC and illustrates the role of vortex acceleration in phonon emission. 
The larger dissipation in oblique collisions seems consistent with the larger acceleration sustained in this case.
On the contrary, the vortex energy dissipated throughout head-on and oblique collisions in UFG and BCS regimes appears equally gradual, being stronger in the latter case. The vortex acceleration in any given configuration should only depend on the absolute value of $d_{in}$ whenever $d_{in}\gg\xi_v$. Thus, the deviating $\Gamma$ measured in the various superfluids signal that acceleration couples to distinct dissipation mechanisms in each regime, adding to the emission of phononic quasiparticles.
One predicted scenario is the emission of fermionic quasiparticles from the vortex cores, as a consequence of the local heating of accelerating vortices \cite{Silaev2012}.

\begin{figure}[b!]
\vspace*{-10pt}
\centering
\includegraphics[width=84mm]{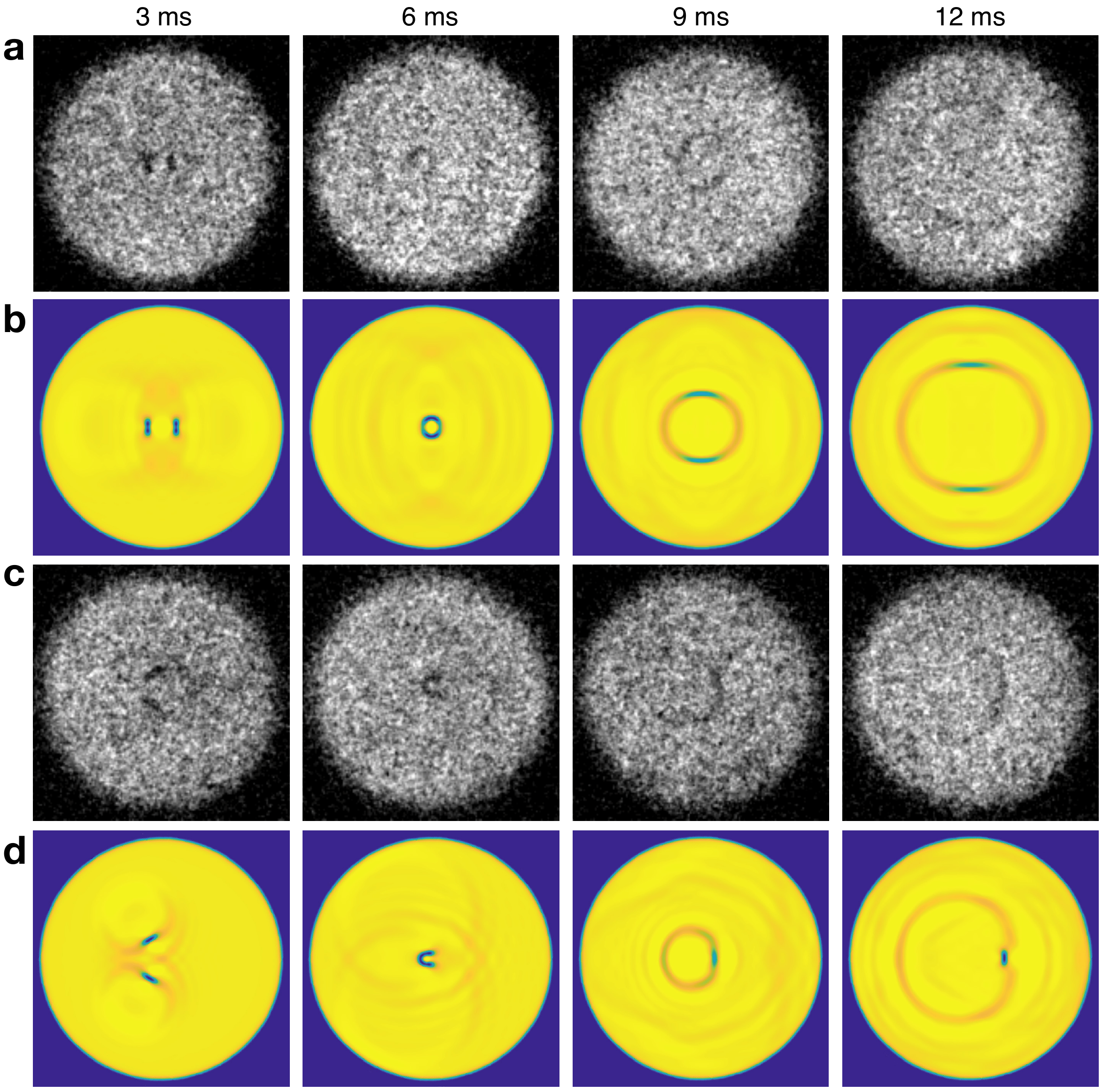}
\caption{\textbf{Direct observation of the sound emission from dipole-dipole annihilation events.} 
Experimental images of annihilation dynamics are shown for \textbf{a}, head-on and \textbf{c}, 120$^\circ$ collisions, recorded in BEC superfluids for $d_{in}< 5\xi$.
GPE simulated images (central $x-y$ plane of the cloud) of the annihilation processes are displayed in panels \textbf{b} and \textbf{d}, for a head-on collision with $d_{in}=2.9 \, \xi$ and a 120$^\circ$ collision with $d_{in}=2.6 \, \xi$, respectively.}
\vspace*{10pt}
\label{Fig4}
\end{figure}

Dipole-dipole collisions offer also a unique opportunity to systematically investigate vortex annihilation. Away from the strong-friction limit, where the self-annihilation of single dipoles can prevail even in the relaxation of superfluid turbulence \cite{kwon2014, Lucas2014, stagg2015, Groszek2016, Baggaley2018}, vortex annihilation is essentially a many-vortex (four-vortex dominant) process \cite{Groszek2016,Cidrim2016,Karl2017,Baggaley2018}.
We determine the annihilation probability $P_a$ by counting the fraction of annihilation events, shown in Figs.~\ref{Fig3}f-h as a function of $d_{in}$.
We estimate the critical initial size $d_a$ for annihilation, such that $P_a (d_{in}\leq d_a) \geq 0.5$, obtaining for head-on collisions $d_a \approx 4 \xi$ in the BEC, $d_a \approx 16/k_F$ in the UFG, and $d_a \approx 19/k_F$ in the BCS regime. In 120$^\circ$ collisions, we mostly observe partial annihilation of the shorter outgoing dipole (Extended Data Fig.~4), occurring at $d_a$ about $\sim 30\%$ larger than the head-on case, because of the final asymmetric configuration \cite{Oblique2019}.
The extracted $\Gamma$ and $d_a$ suggest that annihilation occurs whenever the energy dissipated during the collision is sufficiently large to reduce the initial dipole size to about $d_c$ \cite{Ruostekoski2005}. Here, we exclude that self-annihilations are significant by inspecting the dipole number over time $N_\text{vd}(t)$ (Extended Data Fig.~4).

Annihilation is the most dramatic consequence of vortex-sound interaction, converting both vortex core and incompressible kinetic energies into a sound pulse. We directly visualise the sound emission from dipole-dipole annihilation events both in head-on (Fig.~\ref{Fig4}a) and 120$^\circ$ (Fig.~\ref{Fig4}c) collisions. In the first case, the vortices coalesce shortly after the collision and soon they are converted into a circular density depletion propagating outwards to the boundary of the cloud. The speed of such density pulse is indeed close to the expected sound speed (Extended Data Fig.~5). In 120$^\circ$ collisions, it is difficult to detect sound emission from the partial annihilation of the shorter dipole, due to its small energy. Albeit their very low occurrence probability, complete annihilations are also observed (Fig.~\ref{Fig4}c), where the high-amplitude density pulse travelling rightward is radiated from the annihilation of the longer dipole with higher energy. Such visual evidence of sound emission from vortex annihilations directly discloses the conversion of vortex energy into compressible kinetic energy. 
GPE simulation results (Fig.~\ref{Fig4}b,d) are consistent with experimental observations. Interestingly, we numerically confirm that the density wave created by the head-on collision is not circularly symmetric, since the rarefaction pulse \cite{Jones1982} generated shortly after the annihilation propagates along $y$ axis, and eventually decays into sound (Fig.~\ref{Fig4}b). Moreover, in 120$^\circ$ collisions (Fig.~\ref{Fig4}d), the right-moving dipole does not seem to annihilate permanently, but revives as a U-shaped vortex handle attached to the surface of the condensate. Such low-energy vortex-handle excitations exist in regions of vanishing density, and they are expected to be fragile, thus hardly observable at finite temperatures as in our experiment.
On the other hand, in BCS and UFG superfluids, clear acoustic waves from vortex annihilations are observed rarely (Extended Data Figs.~4,5). This could be associated with the higher speed of sound $c_s$ in these regimes, resulting in low-amplitude density waves since $v/c_s=\delta n/n_0$, where $v$ is the local velocity of atom pairs, and $\delta n$ is the amplitude of the excitation for the density $n_0$. Nonetheless, other possibilities unique of these regimes can be envisioned, such as the total delocalisation of fermionic quasiparticles originally bound in the vortex cores, which may affect sound emission. Our observations are in line with the distinct trends of $\Gamma$, supporting that vortex-sound interactions are not the only relevant dissipation mechanism throughout vortex collisions. %

In conclusion, building a novel platform for 2D vortex collisions in atomic Fermi superfluids, we provide a comprehensive picture of quantum vortex decay arising from mutual friction and vortex-sound interaction, connecting to experiments \cite{Vinen2010} in both superfluid ${}^3$He and ${}^4$He. Most importantly, we visualise vortex annihilation events, providing first direct evidence of the ensuing sound emission. In BECs, we link the vortex energy dissipated into phononic excitations with the acceleration in dipole-dipole collisions, supporting the close relationship between superfluid 2D vortex dynamics and electrodynamics. On the other hand, our observations in BCS and unitary superfluids defy a description solely based on thermal friction and sound emission, pointing to additional contributions from fermionic quasiparticles localised in vortex cores, and from their emission by accelerating vortices. This poses a compelling question about the universal and non-universal aspects of irreversible vortex dynamics \cite{Gabriel2021} and annihilation across the BEC-BCS crossover. 
Extending our work toward 2D quantum turbulence \cite{Simula2014,kwon2016,Gauthier2019,Johnstone2019}, we could study the collisions of clusters, which may further substantiate the specific roles of vortex acceleration in vortex-sound interaction. Another future direction is to investigate vortex-sound interactions from a reversed perspective, letting vortices absorb sound energy \cite{parker2004}. 
The capability of producing programmable vortex configurations in tunable superfluids opens new horizons for vortex research, e.g.~in lower-dimensional quasi-condensates \cite{Karl2017}. Combined with optical disorder potentials,  
this may contribute to the quest of high-performance superconductors by examining exotic phases of vortex matter. 

\bigskip
\begin{acknowledgements}
We thank C.~F.~Barenghi, A.~Bulgac, P.~Magierski, F.~Marino, M.~McNeil Forbes, N.~P.~Proukakis, G.~Wlaz\l{}owski and the Quantum Gases group at LENS for fruitful discussions, and  N.~Cooper for the careful reading of the manuscript.  
This work was supported by the European Research Council under GA no.~307032, the EPSRC under grant no.~EP/R005192/1,
the Italian Ministry of University and Research under the PRIN2017 project CEnTraL, and EU's Horizon 2020 research and innovation programme under Marie Sk\l{}odowska-Curie GA no.~843303.
\end{acknowledgements}

\onecolumngrid



\newpage

\twocolumngrid

\newcommand{\bra}[1]{\mbox{\ensuremath{\langle #1 \vert}}}
\newcommand{\ket}[1]{\mbox{\ensuremath{\vert #1 \rangle}}}
\newcommand{\Li}{$^{6}$Li }
\hyphenation{Fesh-bach}

\newcommand{\bcirc}{\textcolor{blue}{$\bigcirc$}}
\newcommand{\beq}{\begin{equation}}
\newcommand{\eeq}{\end{equation}}

\renewcommand{\theequation}{S.\arabic{equation}}
\setcounter{equation}{0}
\renewcommand{\thetable}{S\arabic{table}}
\setcounter{table}{0}

\setlength{\tabcolsep}{18pt}

\renewcommand{\theequation}{E\arabic{equation}}
\renewcommand{\thefigure}{E\arabic{figure}}
\renewcommand{\thetable}{E\arabic{table}}

\setcounter{equation}{0}
\setcounter{figure}{0}
\setcounter{table}{0}

\setcounter{page}{1}
\begin{center}
\textbf{\large Methods}

\end{center}

\normalsize 

\section{Superfluid sample preparation}\label{Sec:SamplePrep}

\begin{figure*}[t]
\centering
\includegraphics[width=160mm]{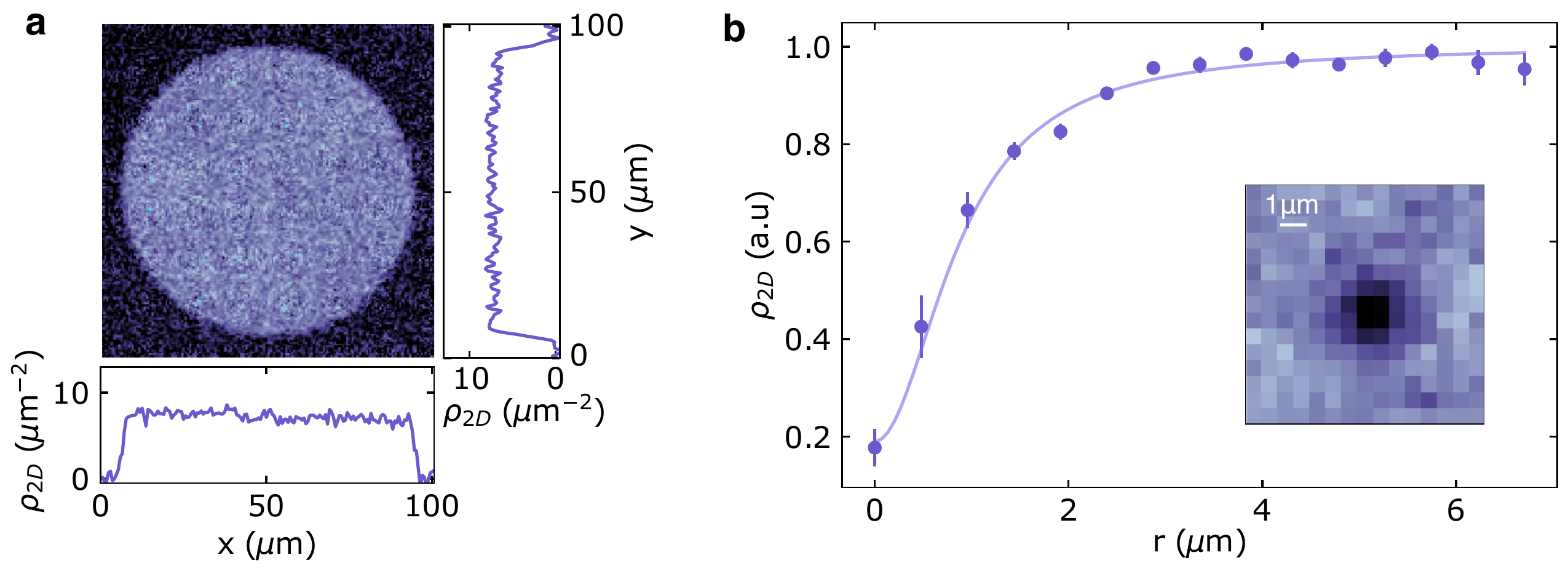}
\caption{\textbf{\textit{In situ} profiles of homogeneous sample and an individual vortex.}
\textbf{a}, In-plane density profile of a UFG sample from a single experimental shot, along with centred vertical and horizontal cuts averaged over 15 different experimental realizations. 
\textbf{b}, \textit{In situ} vortex profile (inset) and its integrated radial profile (symbols) in a BEC sample. The image consists of the average of $10$ experimental realisations. The measured radial profile is fitted with a Lorentzian function (solid line), yielding a width of $0.87(6) \, \mu $m. This matches the expected value $\xi \simeq 0.68\, \mu $m, once the optical resolution of the imaging system is taken into account (see text).}
\label{VortexInSitu}
\end{figure*}

We initially prepare fermionic superfluid samples by evaporating a balanced two-component mixture of the lowest hyperfine spin states of $^6$Li, $\ket{F = 1/2, m_F = \pm 1/2 }$, near their scattering Feshbach resonance at 832\,G in an elongated, elliptic optical dipole trap, formed by horizontally crossing two infrared beams at a $14^{\circ}$ angle \cite{Kwon2020sm,DelPace2021sm}.
A repulsive TEM$_{01}$-like optical potential at 532\,nm with a short waist of about 13\,$\mu$m is then adiabatically ramped up before the end of the evaporation to provide strong vertical confinement. Next, a box-like potential is  turned on to trap the resulting sample in a circular region of the $x-y$ plane (see Sec.~\ref{Sec:Chopstick} and Extended Data Fig.~\ref{VortexInSitu}). 
When both potentials have reached their final configuration, the infrared lasers forming the crossed dipole trap are adiabatically extinguished, completing the transfer into the final uniform pancake trap. Since the box potential has a small dimension with respect to the elongated dipole trap employed for the evaporation, the high-entropy tails of the initial cloud are discarded during the transfer. Within the loading sequence, superfluids at $1/k_Fa \simeq 2.5$ on the BEC and $1/k_Fa \simeq -0.31$ on the BCS side of the crossover are produced by adiabatically sweeping the magnetic field from 832\,G to 857\,G and to 702\,G, respectively. The Feshbach magnetic field coils produce a harmonic confinement in the $x-y$ plane of about 8\,Hz, which is partially cancelled by the repulsive TEM$_{01}$-like potential, yielding a small in-plane trapping frequency of about $2.5$\,Hz. This weak confinement has negligible effect on the sample over the $45 \, \mu$m radius of our box trap, resulting in an essentially homogeneous density \cite{Mukherjee2017sm,Hueck2018sm} (Extended Data Fig. \ref{VortexInSitu}). We further confirm the good uniformity of the cloud by observing the orbiting motion of a long vortex dipole, shown in Extended Data Fig.~\ref{orbit}.

We measure the temperature of the sample before ramping up the in-plane box potential, namely in the composite trap formed by the crossed infrared beams and the green TEM$_{01}$-like beam, by means of usual techniques employed for degenerate gases in harmonic traps \cite{DelPace2021sm}. The temperature measured at unitarity equals $T/T_F  = 0.05(2)$, where $T_F$ is the Fermi temperature given by $k_BT_F=(\hbar k_F)^{2}/2m$. Here, $k_F = (6\pi^2 n)^{1/3}$ is the Fermi wave vector, while $k_B$, $\hbar$ and $m$ are the Boltzmann constant, reduced Planck constant, and the atomic mass of ${}^6$Li, respectively. The very small value of the mutual friction coefficient $\alpha$ measured for a BEC superfluid ensures that the transfer to the homogeneous box potential is adiabatic with no appreciable heating.

The typical final sample consists of about $5 \times 10^4$ atoms per spin state in the box potential with vertical trap frequencies of 356(2)\,Hz, 503(15)\,Hz, and 480(11)\,Hz in the BEC, unitary and BCS regimes, respectively. The resulting ratio between the Thomas-Fermi Radius $R_\text{TF}$ along the vertical $z$-direction, and the characteristic length of the superfluid (characterising also the size of a vortex core) determines the ratio between the vortex line length and core size. For the BEC, unitary and BCS regime, it equals $R_\text{TF}/\xi\approx5$, $R_\text{TF} k_F\approx28$, and $R_\text{TF} k_F\approx31$, respectively. The characteristic lengths in the BEC, unitary and BCS regimes of our samples are $\xi=0.68(2)~\mu $m, $1/k_F=0.27(1)~\mu $m and $1/k_F=0.29(1)~\mu $m, respectively.

\vspace*{-5pt}
\section{Deterministic vortex generation}\label{Sec:Chopstick}

Both the in-plane box potential and the chopstick technique for vortex creation are implemented using a digital micromirror device (DMD), illuminated with blue-detuned 532\,nm light to sculpt arbitrary repulsive optical potentials on the atomic cloud through a high-resolution imaging system \cite{Kwon2020sm, DelPace2021sm}. We obtain dynamical control over the potential by displaying an appropriate sequence of images on the DMD, with a well defined timing between each image set by external triggers. Among several ways to deterministically produce a vortex dipole in BECs \cite{neely2010sm,Aioi2011PRXsm,kwon2015sm,samson2016sm}, the chopstick method \cite{samson2016sm,chopstick_theory2016sm} has the advantage of controlling each vortex, using two focused repulsive obstacles as effective tweezers for a vortex and an antivortex. By acting on the sequence of images displayed on the DMD, this method allows us to adjust the vortex dipole size, position and orientation at will. 

When initially raising the DMD-created potential, the device displays the image of a circular box with an off-centred round obstacle inside. Subsequently, the single obstacle is split into two identical round obstacles, that are moved with a velocity set by the DMD picture-sequence frame rate. Each obstacle is maintained at constant potential height $V_{0}$, even at the initial stage, contrary to Refs.~\cite{samson2016sm,chopstick_theory2016sm} where two independent beams of potential height $V_{0}$ sum up to create an initial obstacle of height $2V_{0}$. At the end of the movement, the obstacles are ramped down by decreasing the number of ON pixels on the DMD in correspondence of their final position. This effectively decreases their height $V_0$, as the finite resolution of the imaging system blurs the discreteness of the DMD image \cite{Kwon2020sm}. 
The DMD-based control of each obstacle is extended towards the creation of many dipoles as in Fig.~2 by employing many obstacle beams from the initial stage. In particular, dipole-dipole collision are engineered by sweeping pairs of obstacles towards the centre of the cloud.

\begin{figure*}[t]
\centering
\includegraphics[width=\textwidth]{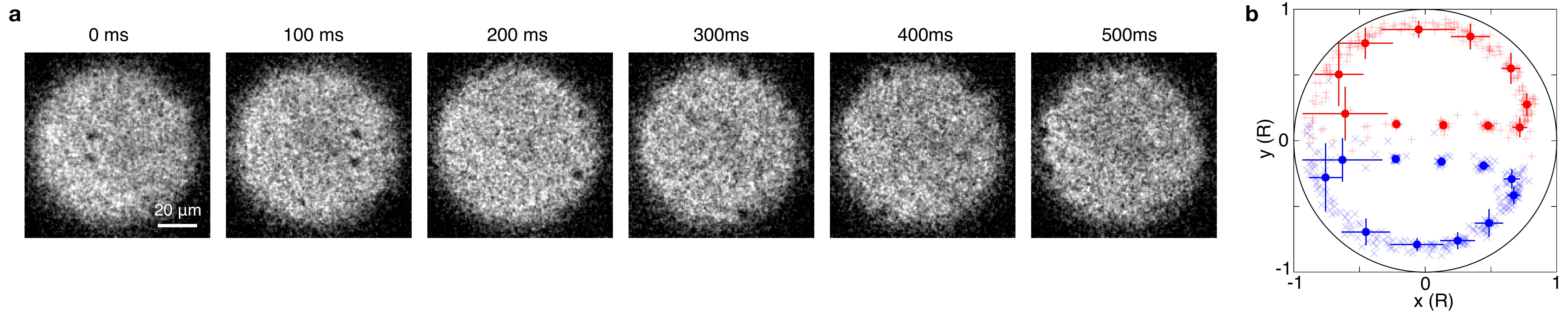}
\caption{\textbf{Orbiting motion of a large vortex dipole in UFGs.} \textbf{a}, A single dipole of $d\approx 12~\mu$m orbiting the homogeneous unitary Fermi gas of radius $R= 45~\mu$m. It rectilinearly crosses the cloud and then orbits it immediately adjacent to the boundary, in stark contrast to a harmonically trapped case where a buoyancy force exists due to the density gradient therein \cite{neely2010sm}.  \textbf{b}, A trajectory obtained from the identical realisations of \textbf{a}. The hold time $t$ varies from $0$ to $500$ ms with time intervals of $50$ ms. The light red $\times$ signs (blue $+$) indicate single realisations of each vortex (antivortex) for the given $t$. The red (blue) circles represent the averaged positions of the vortices (antivortices) at the given $t$. After one orbit $\sim 500-550$ ms, a survival probability of a vortex dipole decreases below $50\%$, 
probably due to interaction with the boundary.}
\label{orbit}
\end{figure*}

We observe that the chopstick protocol works universally well across the BEC-BCS crossover, provided that the speed of the obstacles is appropriately adjusted depending on the interaction strength, yielding a dipole generation probability close to $99\%$. 
Such efficiency does also not depend on the bosonic or fermionic nature of the superfluid, and it is nearly independent of the tuning angle $\theta$ for $10^{\circ}\lesssim\theta\lesssim30^{\circ}$. The shot-to-shot fluctuations of the size $d_{12}$ of the generated dipoles are quantified by an uncertainty of approximately $1 \, \mu$m, corresponding to the standard deviation over more than 40 realisations. 
We attribute such uncertainty to the fact that the size $\sigma=1.3(1)~\mu$m of our obstacles exceeds the vortex core size, allowing a vortex and an antivortex to be located away from the centre of each obstacle.
The chopsticks are optimally designed to be sufficiently small to ensure a stable generation of a short dipole of a few $\xi$, but also sufficiently large to pin and drag each vortex. Correspondingly, we adopt a quite fast ramp-down time of 0.8\,ms of the obstacles, which is necessary to prevent short dipoles from annihilating while obstacles are not yet completely extinguished. On the other hand, for larger dipoles shown in Extended Data Fig.~\ref{orbit}, and doubly-charged dipole generation shown in Fig.~2e, we employ larger obstacles with $\sigma\approx 3.3 \, \mu$m to obtain a more stable dragging and pinning at large $\theta$. We stress that the large obstacle size is indeed favourable to create doubly-charged dipoles.

\vspace*{-10pt}
\section{Vortex Imaging}

In weakly interacting BECs, where the condensed fraction reaches unity at zero temperature, a vortex is signaled by a clear hole in the atomic density over a length scale around the healing length $\xi$ of the condensate. Thus, in the BEC regime we detect vortices by simply employing $1$\,ms time-of-flight (TOF) imaging. However, when working in the strongly interacting regime, a vortex excitation does not produce a significant density-depletion, as the condensed fraction monotonically decreases towards the BCS side of the resonance. To detect vortices here, we therefore perform a rapid sweep of the magnetic field towards the BEC side of the resonance \cite{zwierlein2005vorticessm}, by which the order parameter of fermion pairs can be converted into a BEC of tightly bound molecules. In particular, at unitarity and in the BCS regime, after the hold time $t$, we linearly sweep the magnetic field to 735\,G in $3$ or $3.3$ ms, respectively. The final part of this sweep takes place while the cloud expands in time of flight for 1\,ms. 

Our high resolution imaging system allows also to directly observe vortices \textit{in situ} in the BEC regime \cite{insitu2015sm}. Extended Data Figure~\ref{VortexInSitu} shows the \textit{in situ} image of a vortex (inset) together with its integrated radial profile (main), obtained by averaging over about $10$ experimental realisations.
We extract the vortex core size by performing a Lorentzian fit of the measured radial profile, obtaining a width of $0.87(6)\,\mu$m. We compare this value with the expected width from a numerical simulation of the vortex profile imaged by our setup. For this, we convolve the approximate vortex core profile \cite{pethick2002sm} $\rho(r) \propto r/\sqrt{2\xi^2 + r^2}$ with the independently determined point spread function (PSF) of our imaging system, well approximated by a Gaussian of $1/e^2$ radius of $0.7
\,\mu$m. 
The simulated vortex profile has a Lorentzian width of $0.93(1)\,\mu$m, consistent with the experimental vortex core size. This demonstrates that our estimation of $\xi$ based on the trapping parameters and the atom number is reliable.  
Despite the demonstrated capability to detect vortices \textit{in situ}, we adhere to TOF imaging which benefits from reduced shot-to-shot fluctuations of the vortex visibility.

\section{Data Analysis}
To detect the position of each vortex, we apply an automated sequence that first blurs the experimental absorption image, and then converts it into a binary image, where the density-depleted holes corresponding to vortices appear as particles \cite{kwon2014sm,Cornish2016sm}. 
We extract the position of each particle, and use it as the initial guess of a Gaussian fit of the original image to obtain the position of each corresponding vortex. 
When the automated protocol fails, we estimate the approximate position of each vortex manually, and use this as the initial guess for Gaussian fitting. In this way, we obtain the size of each vortex dipole which is averaged over many realisations (30-100) under the same experimental conditions.

Throughout this work, errors on the mean dipole number $N_{vd}$ are evaluated by adding the standard error of the mean $\sigma_{N}/\sqrt{Z}$ and a measurement uncertainty $1/Z$, where $Z$ denotes the number of measurements, as a quadrature sum $\sqrt{\sigma_{N}^{2}/Z+(1/Z)^{2}}$. The inclusion of $1/Z$ avoids a zero uncertainty for data sets with all equal outcomes.
We note that the mismatch between the values of $d_{in}$ in the $\Gamma$ and $P_a$ data of Fig.~3 arises from the different number of images considered for the estimation of the two quantities. In particular, whenever we measure a non-zero $P_a$, we exclude an equivalent fraction of data in our $\Gamma$ extraction, corresponding to the lowest part of the distribution of $d_{in}$, in order to account for the fraction of annihilation events that are expected to occur for the shortest dipoles.
Thus, to precisely measure $\Gamma$ at very small $d_{in}$ close to annihilation condition, it would be desirable to monitor the collision dynamics in real-time.

\section{Dissipative Point-vortex Model}

A dissipative point-vortex model \cite{Hall1956sm,iordanskii1966sm,schwarz1988sm} is formulated in terms of two mutual friction (dimensionless) coefficients $\alpha$ and $\alpha'$, characterising the longitudinal dissipative and transverse non-dissipative force, respectively. As a first approximation, we only include $\alpha$ in the model as $\alpha'$ is in general quite small \cite{vinen2002Niemelasm}. Taking into account the stationary normal fluid, the model is given by  
\begin{equation}
\mathbf{v_{i}}=\mathbf{v_{i}^{0}}-w_{i}\alpha\, \mathbf{\hat{z}}\times \mathbf{v_{i}^{0}}, 
\label{pointvort}
\end{equation}
where $\mathbf{v_{i}}$ is the resultant velocity of $i$-th vortex, $\mathbf{v_{i}^{0}}$ is the local superfluid velocity created by the other vortices at the position of the $i$-th vortex. The integer $w_{i}$ is the winding number (positive for counter-clockwise flow), and $\mathbf{\hat{z}}$ is the unit vector along the axial (vertical) direction. This model has been successfully employed for describing the dynamics of vortices in atomic BECs in the presence of dissipation \cite{moon2015sm,Kim2016sm,Oliver2020sm}. The velocity of a vortex dipole of size $d$ is $v_d = \hbar/(Md)$, where $M=2m$ is the bosonic (atom pair) mass of ${}^6$Li, is directly verified by the trajectory of single dipoles (Fig.~1e).
By putting this relation into Eq.~(\ref{pointvort}), one can directly obtain the following analytic formula,
\begin{equation}
d(0)^2-d(t)^2=4 \alpha \hbar t/M,
\label{analytic}
\end{equation}
where $d(t)$ is the size of the dipole at time $t$. Consequently, the vortex lifetime $\tau$ is determined by 
\begin{equation}
\tau=\frac{M}{4\alpha\hbar}\left(d(0)^{2}-d_{c}^{2}\right),
\label{life}
\end{equation} 
where $d_c$ denotes the critical length below which a vortex dipole self-annihilates. From these relations, we independently obtain $\alpha$ from the two measurements of Fig.~1f,g, and further the critical dipole length $d_c$ from Fig.~1f.

To decouple the thermal dissipation from the total vortex energy loss in head-on collisions of two dipoles, we numerically solve the equation \eqref{pointvort} inserting the $\alpha$ coefficient measured in each regime (Fig.~1h). We let each dipole start their dynamics at a distance $L= 13 \,\mu$m from the centre of the cloud and evolve until the inter-dipole distance becomes $2L$ after the collision. To satisfy the boundary condition of our circular geometry, we consider the motion of image vortices outside the cloud boundary. We plot the obtained ratio $\Gamma=d_{13} d_{24}/d_{12}d_{34}$ as dotted lines in Fig.~3.

\section{Numerical simulations}


Real-time dynamical simulations of vortex dynamics in the molecular BEC regime at temperature $T=0$ are performed by numerically integrating the mean-field Gross-Pitaevskii equation (GPE)  
\begin{equation}
i \hbar \frac{\partial \Psi}{\partial t} = - \frac{\hbar^2}{2 M} \nabla^2 \Psi +  V \Psi + g | \Psi |^2 \Psi \label{eq.GP.methods}
\end{equation}
for the complex macroscopic wave function $\Psi(x,y,z,t)$,
where $V$ is the external confining potential, $g=4 \pi \hbar^2 a_M/M$ is the strength of the repulsive two-body interaction,
and $a_M=0.6\,a\simeq53.3\,$nm
is the molecular scattering length. 
The external potential $V$ is defined as 
$V(r,z) = \frac{1}{2} M \left ( \omega_\perp^2 r^2 + \omega_z^2 z^2 \right ) + 10 \mu \,\Theta(r/R_0-1)$,
where
$r = (x^2 + y^2)^{1/2}$, and $\{\omega_\perp , \, \omega_z \}= 2\pi \, \{2.5 \, , \, 356\} \text{Hz}$
are respectively the radial and axial trapping frequencies. 
Here, $\Theta(x)$ is the Heaviside step function, $\mu=920.6\,h\times$Hz is the chemical potential, and $R_0=45\,\mu$m is the radius of the cylindrical hard-box potential.
We set the number of particles $N=4.8\times 10^4$, equal to the typical experimental value in BECs, leading to
a healing length $\xi = 0.68\,\mu$m at the centre of the sample.

Our numerical code employs second-order accurate 
finite difference schemes to discretize spatial derivatives;
the integration in time is performed via a fourth-order
Runge-Kutta method. The grid spacings are homogeneous
in the three Cartesian directions 
$\Delta x = \Delta y = \Delta z = 0.4 \, \xi = 0.27 \, \mu$m, 
and the time step $\Delta t = 3.6\,\mu$s. 
The number of grid points in the $x$, $y$, and $z$ direction are
$\{ N_x , N_y , N_z \} = \{ 384 , 384 , 80\}$, 
leading to the computational box 
$-52.82\,\mu {\rm m}\,  \le x \le 52.82\,\mu {\rm m} $, $-\,52.82\,\mu {\rm m}  \le y \le 52.82\,\mu {\rm m} \,$ and $-11.00\,\mu {\rm m} \le z \le 11.00\, \mu {\rm m}$.

\vspace*{-10pt}
\subsubsection*{Vortex imprinting}
We start the simulation with a Thomas--Fermi profile 
for the condensate density $\rho\equiv|\Psi|^2$. In order to calculate the vortex-free ground state, we evolve the GPE in imaginary time until the relative decrease of energy $\Delta E/E$ 
between two consecutive time-steps is smaller than the threshold $\varepsilon=10^{-5}$. 
Once this ground state is reached, we numerically imprint the 
vortices. Vortex imprinting is accomplished by imposing a Pad\'{e} density profile \cite{berloff-2004sm} and a $2\pi$ phase winding around the vortex axis. We then let the system evolve in imaginary time towards the lowest energy state employing the previously described energy convergence criterion. Once $\Delta E/E < \varepsilon$, we introduce a phenomenological dissipation in Eq.~(\ref{eq.GP.methods}) for a very short initial transient to reduce the initial acoustic energy generated by vortex imprinting procedure, and thereafter we start the evolution of the GPE in real time.

\subsubsection*{Vortex tracking}

The algorithm for vortex tracking is based on 
the pseudo-vorticity unit vector 
\begin{equation}
\hat{\bm{\omega}}~:=~\frac{\nabla \Psi_\Re \times \nabla \Psi_\Im}{\left | \nabla \Psi_\Re \times \nabla \Psi_\Im \right |}
\nonumber
\end{equation}
which is tangent to the vortex line along its 
length~\cite{rorai-etal-2016sm,villois-etal-2017sm,Galantucci2019sm}, where
$\Psi=\Psi_\Re + i\Psi_\Im$. 
We reconstruct each $i$-th vortex ($i=1,\dots , 4$) with a spatial resolution $\Delta \zeta = \Delta x/10$. 
At each time $t$ , each vortex is reconstructed as 
$\displaystyle \mathbf{x}_i(t)=\left \{ \left ( x_i(\zeta_j,t) \, , \,  y_i(\zeta_j,t) \, , \, z_i(\zeta_j,t) \right ) \right \}_{j=1,\dots , n_p}$, where $\zeta_j= j \Delta \zeta$ is the discretised arclength and $n_p$ is the number of 
vortex discretisation points. 
Then, in order to determine the trajectories on the $x-y$ plane, we average over the arclength obtaining  
$\displaystyle \langle\mathbf{x}_i(t)\rangle=\left ( \langle x_i(t) \rangle \, , \,  \langle y_i(t) \rangle\, , \, 0 \right )$, as 
vortices are symmetrical with respect to the plane $z=0$.

\subsubsection*{The numerical extraction of the vortex dipole sizes}

Here we describe how we numerically extract the vortex dipole sizes before and after a collision, i.e.~the ratio $\Gamma=d_{13} d_{24}/d_{12}d_{34}$. Since there are tiny dipole size oscillations because of a small amount of sound energy remaining in the condensate, we obtain the dipoles sizes by computing their average value in a certain time interval before and after the interaction period. To identify the latter, we first compute the temporal evolution of the vortex velocity direction $\beta(t)$ defined 
(e.g.~for vortex 1, cf. Fig.~3a) as follows  
\begin{equation}
\beta (t)=\arctan \left(\frac{v_{1,y}(t)}{v_{1,x}(t)} \right) \; \; ,
\end{equation}
where $\displaystyle v_{1,x}(t)=d \langle x_1(t) \rangle/dt$ and 
$\displaystyle v_{1,y}(t)=d \langle y_1(t) \rangle/dt$. Second, we observe that, after an initial constant value,
during the collision the angle $\beta$ changes and, once the interaction between the four vortices ceases, it achieves its post-collision stationary value, indicating indeed a straight trajectory (cf. Fig.~3a). For example, in case of head-on collisions, 
we observe that before the collision $\beta \simeq 0^{\circ}$, it then increases once the interaction with the other dipole starts until  
$\beta \simeq 90^{\circ}$ as the interaction ceases (see Extended Data Fig.~\ref{theory_beta}).  
We thus define the interaction period as $t \in [t_1 , t_2]$, with $t_1$ being the last time instant where $\beta\simeq 0$ 
and $t_2$ the first instant where $\beta \simeq 90^{\circ}$. 
The vortex dipole input size $d_{in}$ is computed by averaging the vortex dipole size from the start of motion until $t_1$, while the vortex dipole output size $d_{out}$ is calculated averaging the dipole sizes from $t=t_2$ until the time
where the inter-dipole distance is equal to $2L$ (in order to mimic the experimental protocol). A similar protocol based on $\beta$ is applied for $120^{\circ}$ collisions. The shaded area in Fig.~3c of the main paper represents the range between $\Gamma+\Delta \Gamma$ and $\Gamma-\Delta \Gamma$, where $\Delta \Gamma$ is a quadrature sum of the standard deviation of $d_{in}$ and $d_{out}$.

\subsubsection*{Dissipation of the incompressible kinetic energy}

The `classical' kinetic energy of a BEC, ${E_k=\frac{1}{2} M\rho \mathbf{v} \cdot \mathbf{v}}$, without considering the quantum energy component $(\hbar ^2/2M) |\nabla \sqrt{\rho}| ^2$, can be decomposed in two terms: the incompressible part $E_k^i$ and the compressible part $E_k^c$. The former includes the energy stemming from vortices, while the latter refers to the acoustic energy of sound waves:
\begin{align*}
\begin{gathered}
E_k ^i = \int \frac{1}{2}M\left [ \left ( \sqrt{\rho} \mathbf{v} \right )^i \right ]^2 d\bold{r},\\
E_k ^c = \int \frac{1}{2}M\left [ \left ( \sqrt{\rho} \mathbf{v} \right )^c \right ]^2 d\bold{r},
\nonumber
\end{gathered}
\end{align*}
where  $\mathbf{\nabla} \cdot (\sqrt{\rho}\mathbf{v}) ^i=0$ and  $\mathbf{\nabla}\times (\sqrt{\rho}\mathbf{v}) ^c=0$, with the fields $(\sqrt{\rho}\mathbf{v}) ^i$ and $(\sqrt{\rho}\mathbf{v}) ^c$  calculated via the Helmholtz decomposition \cite{nore1997sm, numasatosm,horngsm, griffin2019sm,xhaniprlsm,xhaninjpsm}. 
To quantify the amount of incompressible kinetic energy dissipated in sound waves during the collision of two vortex dipoles, we compute the increase of the compressible kinetic energy $\Delta E_k^c$ defined as $\Delta E_k^c=E_k^{c,\rm t_2}-E_k^{c,\rm t_1}$ (see Extended Data Fig.~\ref{theory_ekc}). The dependence of $\Delta E_k^c$ on the input vortex dipole size $d_{in}$ is shown in Fig.~3c (inset) of the main paper. While in Fig.~3c, the dependence of $\Gamma$ on $d_{in}$ is shown only for the cases of collisions without vortex annihilations, the inset includes $\Delta E_k^c$ for smaller values of $d_{in}$ where vortex annihilations could occur.

\onecolumngrid
\clearpage

\begin{figure*}[t]
\centering
\includegraphics[width=90mm]{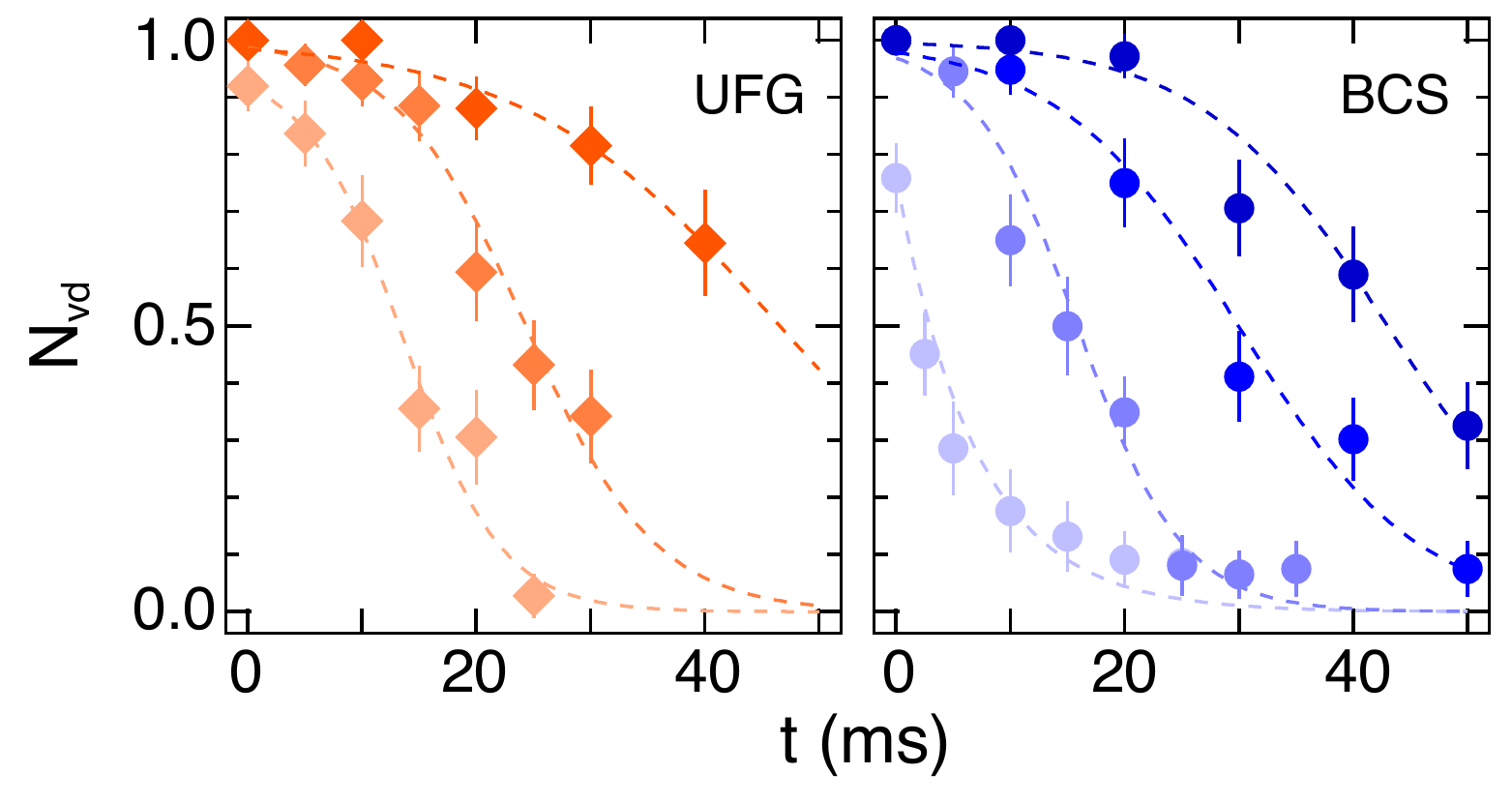}
\caption{\textbf{Decay of short vortex dipoles due to self-annihilation in UFG and BCS regimes.} The dipole half-life $\tau$ for each initial $d_{12}$, i.e.~the time required for $N_{vd}$ to drop to half of its initial value, is determined by fitting $N_{vd}(t)$ with a sigmoid function $1/(1+e^{(t-\tau)/\gamma})$, where the $\gamma$ is used as the measurement uncertainty. The only exception is the shortest dipole shown in the BCS regime, which is fitted with an exponential function. The initial $d_{12}$ is controlled to range from $3.4$ to $6\,\mu$m (lighter colours denotes shorter dipoles). See also Fig.~1f.}
\label{Ext_life}
\end{figure*}

\begin{figure*}[t]
\centering
\includegraphics[width=85mm]{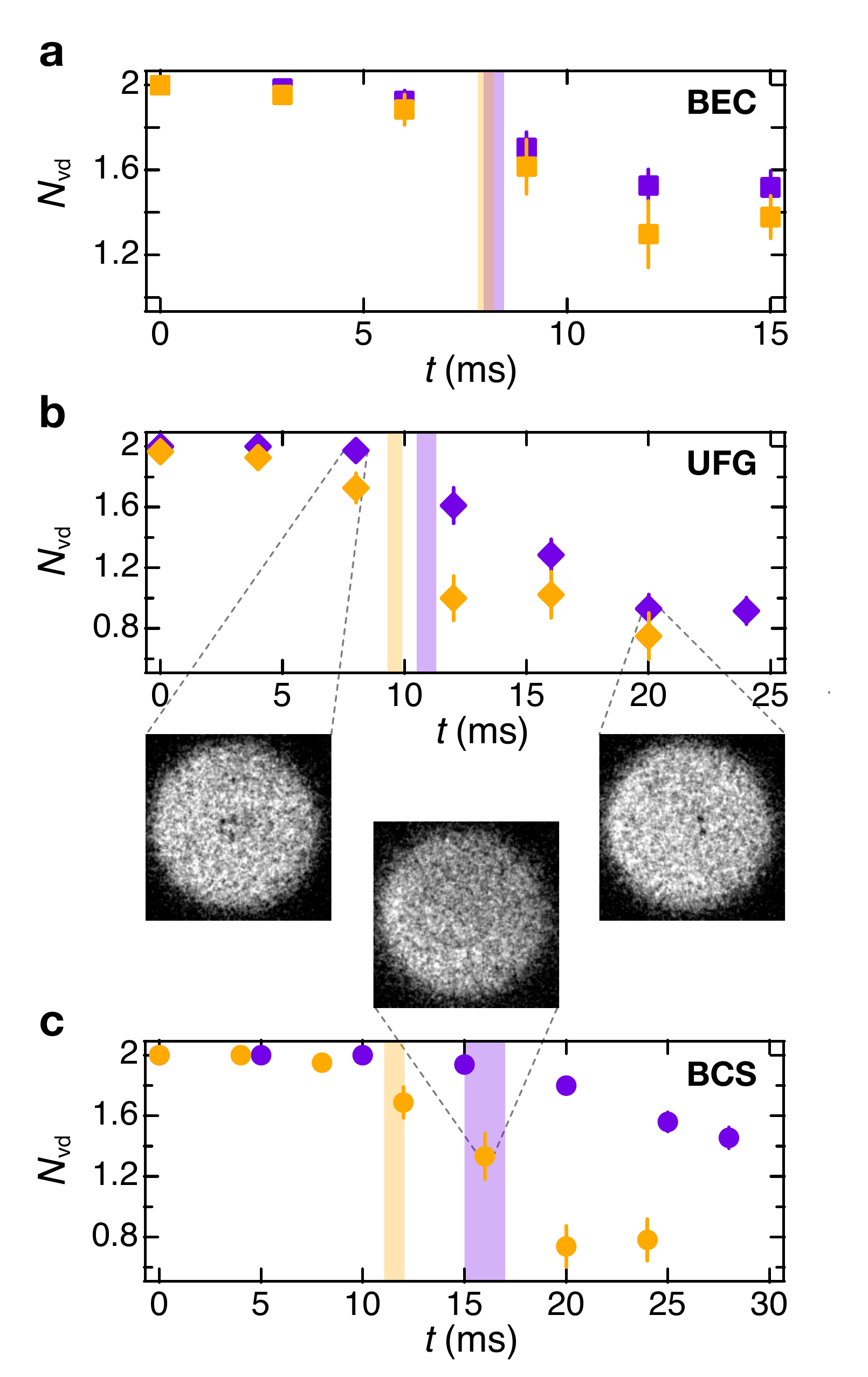}
\caption{
\textbf{Time evolution of the number of vortex dipoles $N_\text{vd}$ during dipole-dipole collisions.} Examples of $N_\text{vd}(t)$ for head-on (120$^\circ$) collisions are shown as orange (purple) symbols in (\textbf{a}) BEC, (\textbf{b}) UFG, and (\textbf{c}) BCS superfluids.
Data sets are part of those for which $P_a$ is shown in Fig.~3f-h of the main text, and specifically: (\textbf{a}) $d_{in} \simeq 5\xi$, (\textbf{b}) $d_{in} \simeq 16/k_F$, and (\textbf{c}) $d_{in} \simeq 18/k_F$ (head-on) and $d_{in} \simeq 24/k_F$ (120$^\circ$).
Shaded regions mark the time interval of vortex partner-exchange during a collision, estimated via DPV model imposing the condition $0.9 < d_{13}(t)/d_{12}(t) < 1.1$. The drop of $N_\text{vd}(t)$ approximately matches this interval, confirming that the observed annihilations do not stem from single-dipole self-annihilations, but are an outcome of the collision dynamics. Experimental images show typical examples of a partial annihilation for a $120^\circ$ collisions (\textbf{b}) and a rarely observed annihilation from head-on collisions in BCS superfluids (\textbf{c}).
}
\label{Ext_Nvd}
\end{figure*}

\begin{figure*}[t]
\centering
\includegraphics[width=110mm]{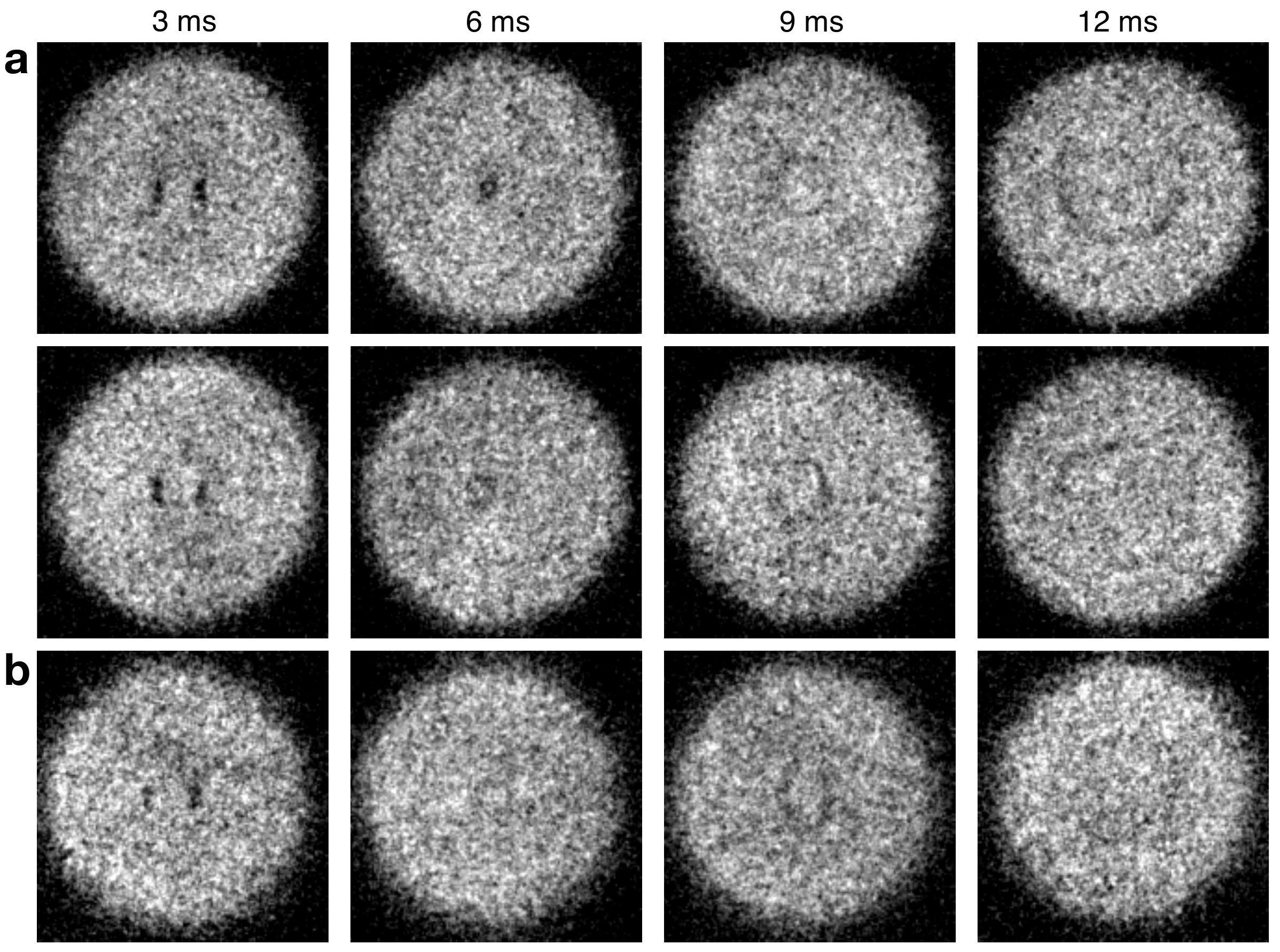}
\caption{\textbf{Annihilation images of BEC and UFGs.} Additional images display the clear emission of a density excitation following vortex annihilations in head-on collisions for (\textbf{a}) BECs and (\textbf{b}) UFGs. Two vortex dipoles collide horizontally as in Fig.~4 of the main text. By roughly measuring the ring sizes of the density pulses observed in BECs ($t=9$\,ms and $t=12$\,ms), we find that the propagation speed of the density pulse is around $4.4$\,mm/s which coincides with the speed of sound evaluated from the density averaged along the tight $z$-direction of the cloud, supporting the acoustic nature of the excitation. Annihilation images observed in UFGs are in general not as clear as in BECs, yet a number of images showing small-amplitude density waves propagating outwards are detected. Each shot is acquired in an independent experimental realisation.}
\label{gallery}
\end{figure*}

\clearpage

\begin{figure*}[t]
\centering
\includegraphics[width=80mm]{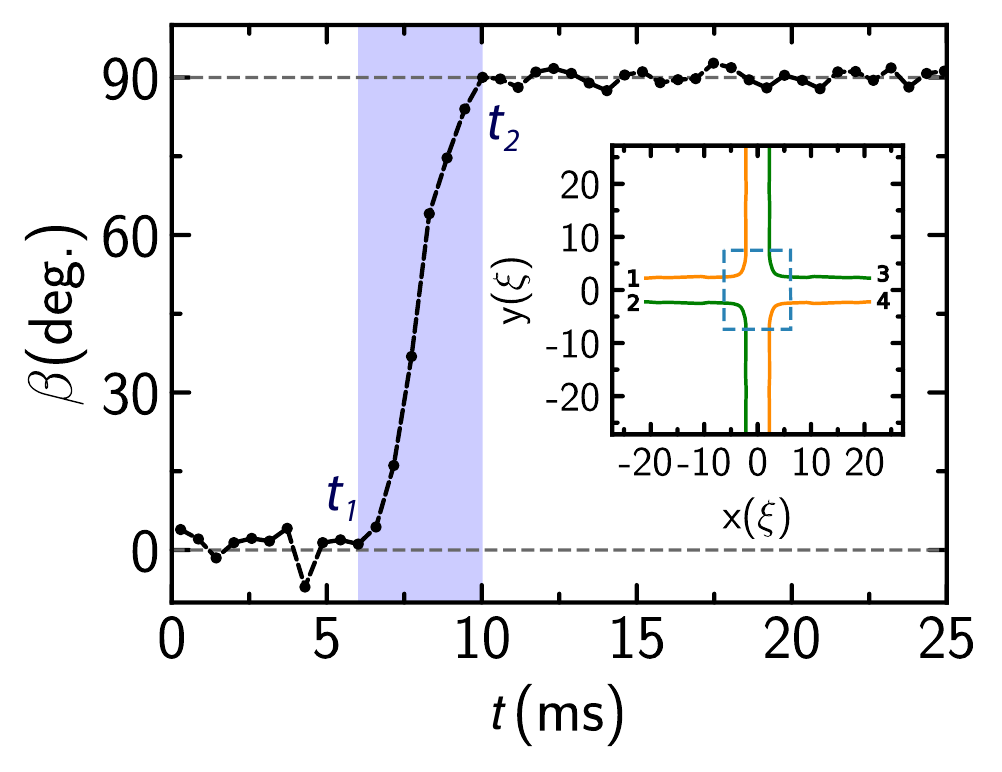}
\caption{\textbf{Numerical criterion for selecting the vortex interaction period.} The temporal evolution of the direction $\beta$ of the velocity of vortex 1 (cf. inset) is displayed for a head-on collision with  
$d_{in}=4.63\xi$. The shaded area indicates the interaction interval $[t_1 , t_2]$ during which the the dipole-dipole interaction takes place, with $t_1$ being the last time instant where $\beta \approx 0^{\circ}$ and $t_2$ the first instant where $\beta \approx 90^{\circ}$. Inset: trajectories of the four vortices in the head-on collision. The dashed blue rectangle denotes the interaction region $[t_1,t_2]$.}
\label{theory_beta}
\end{figure*}

\begin{figure*}[t]
\centering
\includegraphics[width=75mm]{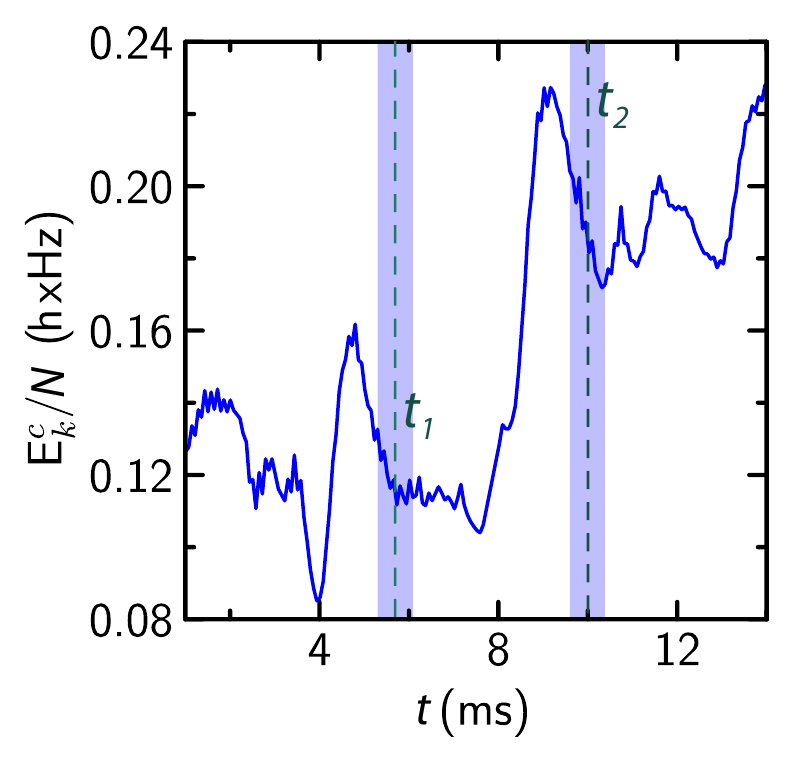}
\caption{\textbf{Time evolution of the compressible kinetic energy $E_k^c$ for the head-on collision and $d_{in}=4.43 \xi$.} The vertical dashed lines indicate times $t_1$ and $t_2$, edges of the interaction interval. The increase of the compressible kinetic energy shown in Fig.~3c (inset) in the main paper is defined as $\Delta E_k^c=E_k^{c,t_2}-E_k^{c,t_1}$. More in detail, the initial $E_k^{c,t_1}$ and the final $E_k^{c,t_2}$ values of the compressible energy are extracted by computing an average value on a 
time interval of width $\delta t$ centred at $t_1$ and $t_2$, respectively.  
This is a characteristic time interval defined as $\delta t=d_{in}/v_{d}$ corresponding to the shaded areas in the plot, where $v_{d}=\hbar/M d_{in}$ is the vortex dipole velocity.
}
\label{theory_ekc}
\end{figure*}

\end{document}